\documentclass[useAMS,usenatbib]{mn2e}

\let\jnlstyle=\rm
\def\refjnl#1{{\jnlstyle#1}}

\def\apjl{\refjnl{ApJ}}                
\def\apjs{\refjnl{ApJS}}               
\def\ao{\refjnl{Appl.~Opt.}}           
\def\aap{\refjnl{A\&A}}                

 \def\mathhMpc{h \mbox{Mpc}^{-1}}

\def\halofit{{\small HALOFIT \ }}          
\def\cubep3m{{\small CUBEP$^3$M}}

\input{psfig.sty}
\usepackage{graphicx}
\usepackage{epsfig}
\usepackage{amssymb}
\usepackage{amsmath}
\usepackage{amsfonts}
\usepackage{txfonts}
\usepackage{multirow}

\title[Baryons, Neutrinos, Feedback and Weak Gravitational Lensing]{Baryons, Neutrinos, Feedback and Weak Gravitational Lensing}
\author[Joachim Harnois-D\'{e}raps et al.]{Joachim Harnois-D\'{e}raps$^{1,2}$\thanks{E-mail: jharno@cita.utoronto.ca}, Ludovic van Waerbeke$^{1}$, Massimo Viola$^{3}$, Catherine Heymans$^{4}$
\\
$^{1}$Department of Physics and Astronomy, University of British Columbia, V6T 1Z1, B.C., Canada\\
$^{2}$Canadian Institute for Theoretical Astrophysics, University of
Toronto, M5S 3H8, On., Canada\\
$^{3}$Leiden Observatory, Leiden University, PO Box 9513, NL-2300 RA Leiden, Netherlands\\
$^{4}$Scottish Universities Physics Alliance, Institute for Astronomy, University of Edinburgh, Royal Observatory, Blackford Hill, Edinburgh, EH9 3HJ, UK\\}

\begin{document}

\date{\today}

\pagerange{\pageref{firstpage}--\pageref{lastpage}} \pubyear{2013}

\maketitle

\label{firstpage}

\begin{abstract}
The effect of baryonic feedback on the dark matter mass distribution is generally considered to be a nuisance to weak gravitational lensing. Measurements of cosmological parameters are affected as feedback alters the cosmic shear signal on angular scales smaller than a few arcminutes. Recent progress on the numerical modelling of baryon physics has shown that this effect could be so large that, rather than being a nuisance, the effect can be constrained with current weak lensing surveys, hence providing an alternative astrophysical insight on one of the most challenging questions of galaxy formation.  In order to perform our analysis, we construct an analytic fitting formula that describes the effect of the baryons on the mass power spectrum. This fitting formula is based on three scenarios of the OWL hydrodynamical simulations. It is specifically calibrated for $z<1.5$, where it models the simulations to an accuracy that is better than $2\%$ for scales $k<10 h\mbox{Mpc}^{-1}$ and better than $5\%$ for  $10  < k < 100 h\mbox{Mpc}^{-1}$. Equipped with this precise tool, this paper  presents the first constraint on baryonic feedback models using gravitational lensing data, from the Canada France Hawaii Telescope Lensing Survey (CFHTLenS). In this analysis, we show that the effect of neutrino mass on the mass power spectrum is degenerate with the baryonic feedback at small angular scales and cannot be ignored.  Assuming a cosmology precision fixed by WMAP9, we find that a universe with no baryon feedback and massless neutrinos is rejected by the CFHTLenS lensing data with 96\% confidence. Some combinations of feedback and neutrino masses are also disfavoured by the data, although it is not yet possible to isolate a unique neutrino mass and feedback model.
Our study shows that ongoing weak gravitational lensing surveys (KiDS, HSC and DES) will offer a unique opportunity to probe the physics of baryons at galactic scales,
in addition to the expected constraints on the total neutrino mass.
\end{abstract}

\begin{keywords}
cosmology: cosmological parameters ---  dark matter --- gravitational lensing: weak lensing --- galaxies: formation --- neutrinos
\end{keywords}


\section{Introduction}
\label{sec:intro}

Recent results from the Canada France Hawaii Telescope Lensing survey (CFHTLenS), a stage II survey \citep{2006astro.ph..9591A}, has demonstrated the power of weak gravitational lensing to probe cosmology \citep{2013MNRAS.431.1547B, 2013MNRAS.429.2249S, 2013MNRAS.430.2200K, 2013MNRAS.432.2433H, 2014MNRAS.442.1326K, 2014arXiv1404.5469F}. While stage III surveys are currently ongoing \citep{2013Msngr.154...44J,2014arXiv1406.4407S}, significant effort is underway in order to reach the precision required by stage IV weak lensing surveys, with the series of GREAT challenges \citep{2014ApJS..212....5M}  devoted to shape measurement, for instance.
A better understanding of the lensing signal at small scales is also necessary and this relies on high resolution numerical simulations. 

It is known that  at small angles smaller (e.g., scales smaller than half a degree, for sources at $z_s \sim 0.5$) the lensing signal  suffers from a large number of theoretical uncertainties: non-linear clustering, projection effects, baryonic physics  to name just a few. On the other hand, the precision of standard big-bang cosmological parameters has improved considerably during the last decade thanks to wide field surveys probing background cosmology \citep[see][and references therein for a review]{2013PhR...530...87W}. The situation today is that our knowledge of most cosmological parameters greatly surpasses our knowledge of the physics of groups and clusters of galaxies. For instance, the mean mass density of the Universe is know to better than $3.5\%$ \citep{2013ApJS..208...19H, 2013arXiv1303.5076P}, which corresponds to a $\sim 6\%$ uncertainty in the mass power spectrum. On the other hand, the uncertainty caused by different Active Galactic Nuclei (AGN) feedback models could be as large as $50\%$  for physical scales $k<1\mathhMpc$ \citep{2014MNRAS.440.2997V}.
This could be particularly problematic for current and future weak lensing surveys since the majority of the signal-to-noise comes from small angular scales \citep{2011MNRAS.417.2020S}, and that in order to measure the dark energy equation of state, it would help considerably to be able to utilize these scales. The other alternative is to restrict lensing analysis to physical scales where these problems are minimized or disappear \citep{2014MNRAS.442.1326K}, or to treat the problematic scales as a nuisance that should be marginalized over \citep{2014arXiv1405.7423E}.

The approach we take in this paper relies on two facts: 1) if all the matter was dark matter, the non-linear clustering would be known very accurately from numerical simulations \citep{2013arXiv1304.7849H, 2014arXiv1406.0543H} and 2) most relevant background cosmological parameters are know to $1-2$ percent \citep{2013ApJS..208...19H, 2013arXiv1303.5076P}. One can therefore assume a fixed cosmology and quantify how strongly the data deviate from the pure dark matter scenario. This deviation can then be compared to various hydrodynamic simulations implementing different models of baryonic feedback, treating the residual uncertainty in the assumed cosmology as a systematic error
in this comparison.

The neutrino mass is the only background cosmology parameter that is not known with great precision and yet, it is very important for our study. \citet{2012PhRvD..85h1101R} and \citet{2013MNRAS.436.2038Z} have measured upper bounds for the neutrino mass, but these studies also show that the modelling at small scales $k > 0.5 \mathhMpc$ is an issue with redshift surveys.
As shown in \citet{2014arXiv1406.5411R}, in the context of the Lyman-Alpha forest, the neutrino mass is best constrained by combining small and large physical scales; this is why gravitational lensing is one of the best approaches for this type of measurement \citep{1999A&A...348...31C}, in particular because the level of modelling at small scales is less complicated than for redshift surveys. Recent attempts at constraining the neutrino mass by combining  CFHTLenS measurements with other cosmology probes suggest that the technique is promising \citep{2014arXiv1403.4599B,2014PhRvL.112e1303B}. Unfortunately, the new developments on the role of AGN feedback show that even with gravitational lensing, baryonic physics has an important effect at small scales. It has been shown recently that the neutrino mass and baryonic feedback are relatively degenerate \citep{2014arXiv1405.6205N}. Our strategy in this paper is therefore to explore the combined effect of different baryonic feedback models and neutrino masses on gravitational lensing measurements, assuming that the background cosmology is known to sufficient accuracy. For this purpose we derive a fitting formula for some specific baryonic feedback models that can be used to predict the matter power spectrum at all scales and redshifts.

In Section \ref{sec:background}, we briefly review the theoretical background relevant for cosmic shear measurements, we describe our different prediction models 
and present a convenient fitting function for different baryon feedback processes. We present the data, the simulations and the measurements in Section \ref{sec:results},
and discuss the results and conclude in Section \ref{sec:discussion}. 
We assume a fiducial cosmology for our simulations and models based on a the WMAP9+BAO+SN $\Lambda$CDM best fit parameters, namely $(\Omega_{\Lambda}, \Omega_M, \Omega_b, n_s, A_s,h) = (0.7095, 0.2905, 0.0473, 0.969, 2.442\times10^{-9}, 0.6898)$. For a zero neutrino mass, the  value of $\sigma_8$ in this fiducial model is $0.831$. 
This number is calculated from $A_s$ for each neutrino mass tested in this analysis. The reason for choosing the WMAP9 cosmology as our baseline (as opposed to a Planck cosmology) roots in a known tension between the Planck and CFHTLenS results \citep{2013arXiv1303.5076P}, 
which could have  biased our analysis towards an over-rejection of theoretical models (see Section for more details).

\section{Background}
\label{sec:background}

\subsection{Theory}

The dark matter power spectrum $P(k)$ is extracted from the dark matter overdensity fields $\delta({\bf x})$ by : 
\begin{eqnarray}
\langle | \delta ({\bf k}) \delta ({\bf k'}) | \rangle = (2\pi)^{3}P({\bf k})\delta^3_{D}({\bf  k'} - {\bf k})
\label{eq:power}
\end{eqnarray}
where $\delta ({\bf k})$ is  the Fourier transform of $\delta({\bf x})$, and $P(k)$ is obtained by averaging $P({\bf k})$ over all directions.
Under the Limber approximation \citep{1954ApJ...119..655L},  the weak lensing power spectrum $C_{\ell}^{\kappa}$ is related to matter power spectrum with:
\begin{eqnarray}
C_{\ell}^{\kappa} =  \frac{1}{\ell}\int_{0}^{\infty} dk W^2(\ell/k) P(k,z) \mbox{, \hspace{0.2cm}} W(\chi) = \frac{3 H_{0}^{2} \Omega_{M}}{2 c^2}\chi g(\chi)  (1 + z)
\label{eq:limber}
\end{eqnarray}
where $\ell = \chi k$, $c$ is the speed of light, $H_0$ the Hubble parameter, $\Omega_{M}$ the mean matter density in units of critical density, $\chi$ the comoving distance to the observer and
$g(\chi)$ describes the lensing geometry of the system,  with a source redshift distribution $n(z)$:
 \begin{eqnarray}
         g(\chi) =  \int_{\chi}^{\chi_H} n(\chi') \frac{\chi' - \chi}{\chi'} d\chi' 
\end{eqnarray}
The cosmic shear correlation functions $\xi_{\pm}(\theta)$ are computed from this quantity with:
\begin{eqnarray}
    \xi_{\pm}(\theta) = \frac{1}{2\pi}\!\int\!\!\!C_{\ell}^{\kappa} J_{0/4}(\ell\theta)    \  \ell \ d\ell 
     \label{eq:xi}
\end{eqnarray}
where $J_{0/4}(x)$ are  Bessel functions.

\subsection{Models of $P(k)$}
\label{subsec:models}

\subsubsection{Dark matter only models}

Although the largest scales of the matter field can be accurately described by linear perturbation theory, 
the smallest scales require modelling of the non-linear regime of gravitational collapse.
The weak lensing measurements we analyze in this work extend down to sub-arcminute scales, hence
it is necessary to include scales up to $k=40\mathhMpc$ in the model predictions.
These are very deep into the non-linear regime, where the modelling is not fully tested, thereby it is essential to quantify the theoretical uncertainties.
Our approach is to assume a fixed, fiducial cosmology, 
compare a series of theoretical predictions for $P(k)$, and estimate the error on the theory 
as the scatter across the models (see  Table \ref{table:models} for a list of the models considered in this work).

{\it HF2 model: } The power spectrum from the widely used \halofit \citep{2003MNRAS.341.1311S}  fitting function including its recalibration by \citet{2012ApJ...761..152T}.  
Known limitations include 5-10 percent over-prediction of power for $0.5 < k < 5 \mathhMpc$ in standard $\Lambda$CDM cosmology \citep{2010ApJ...715..104H},
mainly due to a coarse sampling of the cosmological parameter space. Smaller scales deviate from other models, hence we consider this model to be 10 percent accurate.

{\it CEHF model: } An alternative to universal fitting functions  has been proposed by \citet{2010ApJ...715..104H}, which instead 
interpolate the power spectrum from an ensemble of well-controlled $N$-body simulations.  
This Cosmic Emulator has been shown to be accurate at the percent level up to $k = 1\mathhMpc$ and 
5 percent up to $k = 10 \mathhMpc$ \citep{2013arXiv1304.7849H}. Smaller scales are not available with the Cosmic Emulator, 
which is unfortunate for weak lensing studies since these scales contribute significantly to the shear correlation functions at the arcminute level
\citep{2014arXiv1406.0543H}. 
Following \citet{2011MNRAS.418..536E}, we extend the Cosmic Emulator at smaller scales by 
grafting the HF2 predictions, with an overall normalization factor to ensure continuity across  the junction.
The grafted scales are considered to be 10 percent accurate.

{\it CEp model: } By construction, the CEHF model reproduces the same shape as HF2 at small scales, 
which is not guaranteed to be accurate. We therefore devise another empirical model in which the 
Cosmic Emulator is extrapolated to smaller scales by a simple power law, fit over the range $5<k< 10 \mathhMpc$, 
and  then extended to $k=40\mathhMpc$.
When compared with high resolution simulations (see the HR model below),  we find that a function of the form
\begin{eqnarray}
 P^{CEp}(k) \propto k^{\alpha(z) - 3.0} \mbox{, \hspace{1cm}} k>10\mathhMpc
\end{eqnarray}
with $\alpha(z) = 0.92(1+z)^{0.1}$ provides a smooth and precise extrapolation for $z\le2$.
The proportionality constant is simply found by matching the amplitude at $k=10 \mathhMpc$.
We note that higher redshifts are better described with higher values for $\alpha$ than those prescribed here (up to 25 percent higher by $z=3$).
Given the redshift distribution for our sample of  CFHTLenS galaxies has a mean redshift  $<z_s>\sim 0.9$ (see Section \ref{subsec:data}), this correction on $\alpha$ has negligible effect on our measurement.
Compared to CEHF, this model has the extra advantage that its derivative is continuous, which is desired for most Fisher matrix calculations. 
The accuracy of the grafted scales are taken to be 10 percent up to $k=20\mathhMpc$, and 20 percent for smaller scales, to be conservative. 

{\it HF1b model: } Before the recalibration by \citet{2012ApJ...761..152T}, the \halofit model (HF1) was under predicting the small scale power by 
up to a factor of two  \citep{2010ApJ...715..104H}.    
However, an analytical rescaling of the original \citep{2003MNRAS.341.1311S}  predictions, proposed by John Peacock \footnote{\tt www.roe.ac.uk/$\sim$jap/haloes},
 was found  to reproduce with high fidelity the results from high resolution $N$-body simulations. 
 This model can be constructed from \halofit ($<2012$ versions) as:
\begin{eqnarray}
P^{HF1b}(k) = \bigg[ P^{HF1}(k) - P_{lin}(k) \bigg]   \times \frac {1+2y^2}{1+y^2} +  P_{lin}(k)
\label{eq:corrfactor}
\end{eqnarray}
where $y = k/(10 \mathhMpc)$. 
As a result, HF1b is considered in this work, but not HF1, and the accuracy is taken to be 10 percent for $0.1<k<1\mathhMpc$, 15 percent for $k>1\mathhMpc$, and 5 percent at smaller $k$-modes.


{\it HR model:} Our last candidate for $P(k)$ is taken from a recent  $N$-body simulation suite, 
the Scinet LIght Cone Simulations (SLICS), which achieve a precision better than five percent  for scales of $k<30 \mathhMpc$ in the high resolution series.
Features of the SLICS series are summarized in Section \ref{subsec:sims} and detailed in \citet[][HDVW hereafter]{2014arXiv1406.0543H}. 
This is not a model {\it per se} but a measurement estimated from light cone simulations created with an independent $N$-body code; 
it is therefore an important indicator of the level 
of precision that is achievable. We treat the HR model as an additional estimate of the signal, with five percent accuracy  to $k<10 \mathhMpc$,
10 percent accuracy to $k=20 \mathhMpc$ and 20  percent accuracy to $k=30\mathhMpc$. 
For $k>30\mathhMpc$, the model is  considered to be precise to within a factor of two, 
effectively downweighting the regions that suffer from limitations due to mass resolution in the $N$-body calculation.

\begin{figure}
   \centering
    \includegraphics[width=2.9in]{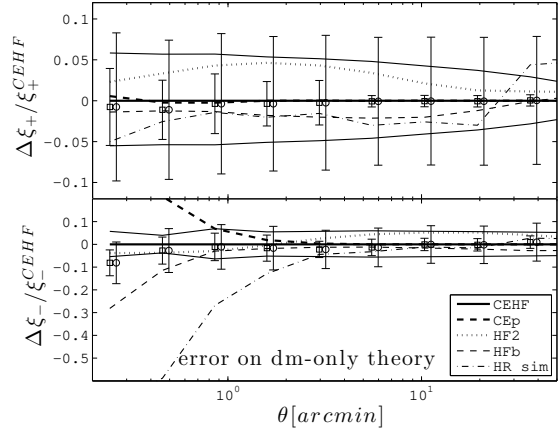} 
      \caption{Fractional error on the dark matter only theoretical models for shear correlation functions $\xi_+$ (top) and $\xi_-$ (bottom). 
      Results are compared to the CEHF model. Squares with error bars are the weighted mean and error across the different models (see the main text for details about the variance on individual models).
      The upper  (lower) thin solid lines in each panel correspond to the effect of a $1\sigma$ upward (downward) fluctuation in $\Omega_{M}$ on the CEHF model, compared to the baseline cosmology.  
      The open circles (slightly shifted for clarity) represent the same weighted mean, but the larger error bars combine in quadrature the theoretical error on the dark matter model
       and  the uncertainty on $\Omega_{M}$. 
      }
   \label{fig:model_err}
\end{figure}

For each of these dark-matter only models, we compute the shear correlation functions $\xi_{\pm}$ and report the result in Fig. \ref{fig:model_err},
organized as fractional difference with respect to the CEHF model.
The agreement between the different $\xi_+$ models is at the level of a few percent even at 0.3 arcminutes.
We find HF2 to be the main outlier.
Models of $\xi-$ achieve the same level of accuracy down to about 3 arcminutes, but smaller angles
 do not reach the same level of agreement. 
  The squares with error bars in Fig. \ref{fig:model_err} are the weighted mean and  error ($1\sigma$) across models,
 obtained by weighting each model by its inverse variance. Overall, we achieve a  1 percent precision on $\xi_+$ for angles larger than 5', 
 and a 4 percent precision for smaller angles. The precision on $\xi_-$ is poorer as this quantity is probing deeper into the non-linear regime:
 we achieve four percent precision on angles larger than 3', and an eight percent precision for smaller angles. 
 Recall that this is the error on the non-linear weak lensing signal for a fixed cosmology universe in which there is no baryonic feedback nor massive neutrinos. 
  
  We also show in the figure the effect on the CEHF model of a $1\sigma$ fluctuation in $\Omega_M$ compared to the fiducial value.
  The open circles with larger error bars show the combined (model + cosmology) uncertainty, as fully described in Section \ref{subsubsec:cosmo_err}.

\subsubsection{Neutrino feedback}

The effect of the neutrino free-streaming on the dark matter structure has been calculated from simulations with a high level of precision
and incorporated into the {\small CAMB } cosmological code \citep{Lewis:1999bs, 2012MNRAS.420.2551B} with less than 10 percent error at $k = 10 \mathhMpc$. 
With this tool, we compute the  mass power spectrum for our model with both dark matter and massive neutrinos, $P_{HF2}^{DM + \nu}(k)$, assuming one massive and two massless flavours. We explore three different total neutrino masses $M_{\nu}$ of $0.2$, $0.4$, and $0.6$ eV in addition to the massless case $M_{\nu}=0$. The ratio between these and the dark matter only model provide 
our four predictions of the {\it neutrino feedback bias}: 
\begin{eqnarray}
b^2_{M_{\nu}}(k,z) \equiv \frac{P^{DM+M_{\nu}}(k,z)}{P^{DM}(k,z)},
\label{eq:nu_feedback}
\end{eqnarray}
where the $M_{\nu}$ superscript specifies the total neutrino mass considered.
For each model $X$ of Table \ref{table:models}, we implement the neutrino feedback with a multiplicative bias factor, i.e.
$P^{DM + \nu}_{X} = P^{DM}_{X} \times b^2_{M_{\nu}}$, with $X=$ (HF2, HF1b, ...).


\subsubsection{Baryon feedback}


The baryonic feedback models are obtained\footnote{OWL simulations: {\tt http://vd11.strw.leidenuniv.nl/}} from a subset of the hydrodynamical simulation suite ran in the context of the OverWhelmingly Large (OWL) Simulation Project \citep{2010MNRAS.402.1536S}. The dark matter density fields of these simulations were compared to a dark matter only baseline, and discrepancies were reported as 
baryonic feedback on the dark matter \citep{2011MNRAS.415.3649V}. Amongst different models, we selected four models: 1) the dark matter only (DM-ONLY) 2) the reference baryonic 
model (REF) that describes prescriptions for cooling, heating, star formation and evolution, chemical enrichment and supernovae feedback and 3) a model that has an additional contribution from the active galactic nuclei feedback (AGN), and 4)  a top-heavy stellar initial mass function (DBLIM), but no AGN feedback \citep[see][for details about these simulations]{2011MNRAS.415.3649V}.
Following \citet{2011MNRAS.415.3649V, 2011MNRAS.417.2020S}, we model the baryonic feedback on dark matter by taking the ratio with the DM-ONLY model, 
and define the {\it baryon feedback bias} as:
\begin{eqnarray}
   b^2_m(k,z) \equiv \frac{P^{DM+b(m)}_{\rm OWL}(k,z)}{P^{DM}_{\rm OWL}(k,z)},
   \label{eq:baryon_bias}
\end{eqnarray}
where the index $b(m)$ runs over the different baryon feedback models (AGN, REF, ...), and the subscript OWL specifies that these quantities are measured from the OWL simulation suite. 
The lower section of Table \ref{table:models} summarizes the baryonic feedback models considered in this paper.

\subsubsection{Combined feedback}

In this analysis, we consider all combinations of the four neutrino masses  (three with  $M_{\nu} > 0$, plus the massless case)  with the four baryon feedback models (three with baryonic physics,  plus  the no baryon case)
for a total of 16 models, all constructed from:
\begin{eqnarray}
P^{DM+\nu+b(m)}(k,z) = P^{DM}(k,z) \times b^2_m(k,z)  \times  b^2_{M_{\nu}}(k,z)
\end{eqnarray}
The underlying assumption from this `multiplicative' parameterization is that  the baryonic feedback is independent of the  neutrino free streaming.
This statement is justified since \citet{2012MNRAS.420.2551B} found that baryons have a one percent effect on the neutrinos
for $k<8 \mathhMpc$ with a gradual increase at smaller scales. This is clearly subdominant compared to the baryon feedback itself, reinforcing the validity of our multiplicative feedback method. 

\begin{figure}
   \centering
   \includegraphics[width=3.4in]{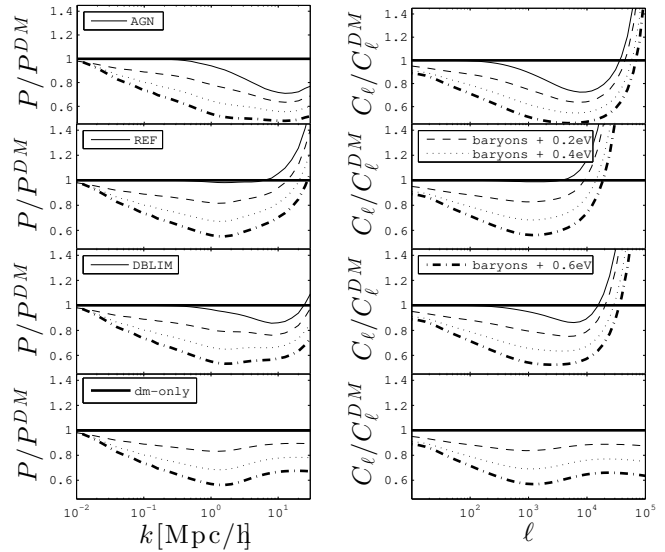} 
      \caption{(Left:) Combined feedback from baryons and massive neutrinos on the  dark matter power spectrum, measured at $z = 0.042$. 
     Each panel shows the dark matter only model as the thick horizontal line, and the dark matter + baryons as the thin solid line. 
      Top to bottom are  AGN, REF,  DBLIM and DM-ONLY baryon models respectively. 
      Also shown is the impact of neutrinos on each model,
      shown as thin dashed lines ($0.2$eV), dotted lines ($0.4$eV), and thick dash-dotted lines ($0.6$eV). 
      (Right:) Same as the left panel, but for the weak lensing power spectra, assuming the  source redshift distribution given by equation \ref{eq:nz}.}
   \label{fig:PkCl}
\end{figure}

The left panels of Fig. \ref{fig:PkCl} illustrate the action of different combinations  of baryons and massive neutrinos on the dark matter power spectrum.
The right panels show the same combinations propagated on the weak lensing power spectrum $C_{\ell}^{\kappa}$ with equation \ref{eq:limber}. 
As noted by \citet{2014arXiv1405.6205N}, the two sources of feedback are highly degenerate for $\ell>1000$ and will be challenging to distinguish in coming surveys.
 The region with  $\ell<1000$ is more sensitive to neutrino masses and could break the degeneracy, although it is more affected by sampling variance.
 The optimal choice will be affected by the mean source redshift and the noise level, and will therefore differ slightly in each survey.

\begin{table}
   \centering
   \caption{The theoretical models considered in this paper. 
   The Cosmic Emulator (CE) has a small scale $k$-cut at $10.0 \mathhMpc$, which affects many scales relevant for the current studies.
    As described in the main text, we therefore extend the CE to smaller scales by grafting either the {\small HALOFIT}2012 predictions (CEHF model)
    or  a power law (CEp model). References for these models are also provided in the main text.}
      \begin{tabular}{@{} ll ll } 
      \hline
     Description&   $k$-modes included  & Name\\
             & [ in $h \mbox{Mpc}^{-1}$]  & \\
\hline
           {\small HALOFIT}2012                           &     $0.001 < k < 40.0$                 &  HF2\\   
           {\small HALOFIT}2011 `corrected'          &    $0.001 < k < 40.0$                 &  HF1b\\
               Cosmic Emulator               &     $ 0.001 <  k < 10.0 $              & CEHF\\
                     $+$ \halofit2012  extension                                     & $ 10.0 <  k < 40.0 $ &   \\  
           Cosmic Emulator               &       $ 0.001 <  k < 10.0 $         & CEp  \\
               +      Power law    extension                                 & $ 10.0 <  k < 40.0 $ & \\      \hline
             Large ensemble suite  ($N_{sim}=500$)& $ 0.0124 <  k < 20.0 $ & SLICS-LE \\
             High resolution  suite ($N_{sim}=5$) & $ 0.0124 <  k < 20.0 $ & SLICS-HR \\ \hline
             Dark matter baseline for baryons& $ 0.0013 <  k < 100.0 $ & dm-only\\
             Reference for baryonic feedback & $ 0.0013 <  k < 100.0 $ & REF \\
             REF + AGN   feedback      & $ 0.0013 <  k < 100.0 $ & AGN \\
            REF + top-heavy IMF & $ 0.0013 <  k < 100.0 $ & DBLIM    \\ \hline
   \end{tabular}
    \label{table:models}
\end{table}

\subsection{Fitting formula for baryon feedback  }
\label{subsec:fit}

Whilst massive neutrinos are already featured in {\small CAMB}'s  mass power spectrum predictions, baryon feedback, however, is not included.  
For this purpose, we provide a fitting function for the three baryonic feedback models considered here (REF, DBLIM and AGN). For each model, our fitting function is designed to reproduce the baryonic effects on the total mass power spectrum for any redshift and  scale  with high precision. It can then easily be incorporated to {\small CAMB} or any other tool to create fully non-linear power spectrum predictions that include baryonic effects.

\begin{figure}
   \centering
   \includegraphics[width=2.9in]{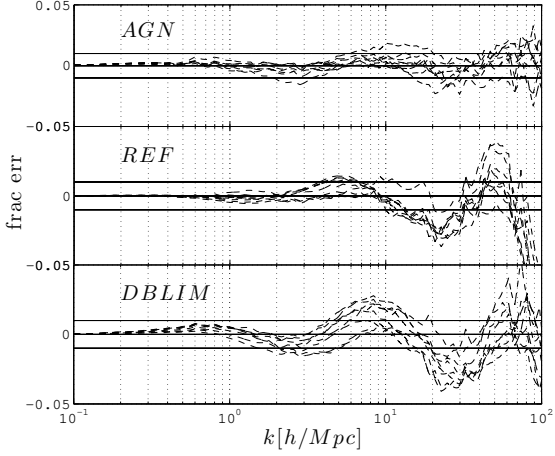} 
      \caption{Fractional error between the baryon bias fit function (equation \ref{eq:fit_baryon}) and that measured by \citet{2011MNRAS.415.3649V}. 
       The horizontal lines highlight the 1 percent error. Each panel contains the result for 8 different redshifts in the range $0 \le z \le 1.5 $. 
       Higher redshifts show stronger errors, as mentioned in the text. Nevertheless, the fit still describes the bias with sub percent level precision up to  $k\sim 1 \mathhMpc$ 
       and better than 10 percent precision up to $k\sim 20 \mathhMpc$. Smaller scales contribute negligibly to the cosmic shear signal, unless probing deep 
       in the sub-arcminute regime.}
   \label{fig:fit_err}
\end{figure}

For the three feedback models considered here, we find that the baryon feedback bias is well described by the following functional form:
\begin{eqnarray}
b^2_m(k,z) = 1-A_{z}e^{(B_{z}x-C_{z})^3}+D_{z}x e^{E_{z}x}
\label{eq:fit_baryon}
\end{eqnarray}
with $x = \mbox{log}_{10}(k/[\mathhMpc])$.
The five terms $A_z$, $B_z$, $C_z$, $D_z$ and $E_z$ depend on redshift, 
closely following  a quadratic polynomial in powers of the scale factor, i.e. 
$A_{z} = A_{2} a^2 + A_{1} a + A_{0}$, with  $a = 1/(1+z)$. The best fit parameters for each model are presented in Table \ref{table:fit_baryon},
and compared to a direct interpolation from the measurements of \citet{2011MNRAS.415.3649V} in Fig. \ref {fig:fit_err}, for $z<1.5$.
We see that this parameterization is accurate at the sub percent level for  $ k < 1 \mathhMpc$ and 
the fractional error is generally less than 5 percent even for $k \sim 100 \mathhMpc$. 
At higher redshift, the fit is still good but shows stronger discrepancies with the interpolation method:
in all models, scales and redshifts, the error never exceeds 15 percent for $k < 40 \mathhMpc$, 
or 33 percent for $k = 100 \mathhMpc$.
The fitting formula is accurate to (1, 5, 10) percent at  $k =$ (0.7, 1.5, 20, AGN).
($1.5$, $25$, $35$, REF) and ($1.0$, $15$, $30$, DBLIM), where $k$ is given in  $\mathhMpc$.

\begin{table}
   \centering
   \caption{Best fit parameters that describe the baryonic feedback on the matter power spectrum extracted from the OWL simulations.
   Given a model $m$ (AGN, REF or DBLIM) and a scale factor $a = 1/(1+z)$, this Table allows the reconstruction of  
   the five terms that enter the baryon feedback bias $b_{m}(k,z)$ (equations \ref{eq:baryon_bias} and \ref{eq:fit_baryon}).
      The index $i$ refers to the power of $a$ associated with the coefficient. 
   For example, the first term is constructed as  $A_{z} = A_{2} a^2 + A_{1} a + A_{0}$.}
      \begin{tabular}{|l|l|cccccccccc|} 
      \hline
     $m$  &$i$&   $A_{i}$& $B_{i}$& $C_{i}$& $D_{i}$ & $E_{i}$ \\
         \hline
         \multirow{3}{*}{AGN} 
    &2  &  -0.119  &0.130  &0.600  &0.00211 &-2.06\\
    &1  &  0.308 &-0.660 &-0.760 &-0.00295  &1.84\\
    &0  &  0.150  &1.22   & 1.38    & 0.00130  &3.57 \\
     \hline
      \multirow{3}{*}{REF} 
  &2&   -0.0588  &-0.251  &-0.934 &-0.00454   & 0.858\\
  &1&      0.0728 &  0.0381& 1.06  &  0.00652  & -1.79\\
  &0&     0.00972 &1.12    &  0.750   &-0.000196 & 4.54\\
      \hline
     \multirow{3}{*}{DBLIM} 
     &2& -0.295  & -0.989 &-0.0143  &0.00199 & -0.825\\
    &1&   0.490   & 0.642 &-0.0594 &-0.00235 & -0.0611\\
     &0&  -0.0166& 1.05    &1.30     & 0.00120   & 4.48\\
      \hline
      \end{tabular}
    \label{table:fit_baryon}
\end{table}
 
\begin{figure}
   \centering
   \hspace{2cm}
      \includegraphics[width=2.7in]{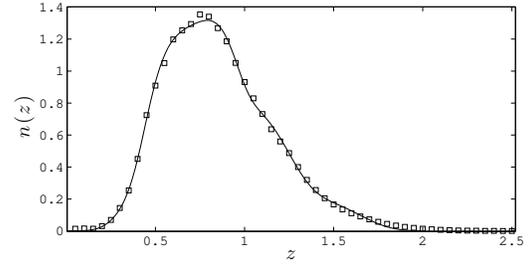} 
      \caption{Redshift distribution from CFHTLenS (black squares) corresponding to galaxies with $0.4<z_{\rm phot}<1.3$, 
     with an arbitrary normalization. The solid line is the best fit given by Eq. \ref{eq:nz}.}
   \label{fig:fit_zPDF}
\end{figure}

\section{Results}
\label{sec:results}

\subsection{Data}
\label{subsec:data}

We use the public release of the Canada France Hawaii Telescope Lensing Survey (CFHTLenS\footnote{CFHTLenS: {\tt www.cfhtlens.org}}) to measure the shear correlation functions $\xi_{\pm}$. CFHTLenS spans a total survey area of 154 deg$^2$, constructed from a mosaic of 171 individual pointings observed by the 1 deg$^2$ imager at the Canada France Hawaii Telescope. The survey consists of four compact regions called W1, W2, W3 and W4, which cover approximately 64, 23, 44 and 23deg$^2$ respectively. Details on the data reduction are described in \citet{2013MNRAS.433.2545E}. The effective area is reduced to 120 deg$^2$ by the masking of bright stars, artificial and natural moving objects and faulty CCD rows. The observations in the five bands $u'griz$â of the survey allow for the precise measurement of photometric redshifts \citep{2012MNRAS.421.2355H}. The shear measurement with {\it lens}fit is described in detail in \citet{2013MNRAS.429.2858M}. The residual systematics for galaxy shapes are described in \citet{2012MNRAS.427..146H} and the reliability of photometric redshifts is quantified in \citet{2013MNRAS.431.1547B}.

As described in  \citet{2012MNRAS.427..146H}, the star-galaxy shape cross-correlation is the objective criteria, insensitive to cosmology, that is used to flag an individual pointing as {\it good} or {\it bad}, depending on its probability to be contaminated by residual Point Spread Function (PSF) distortions. 
In addition, weak multiplicative and additive shear calibration factors $m$ and $c$ are calculated and applied; 
in this work, we revisit the 
$c$-correction in order to make it less dependent on an arbitrary parametric model. In \citet{2012MNRAS.427..146H}, the $c$-correction is modelled as a function of the galaxy signal-to-noise $\nu_{\rm SN}$ and size $r$, finding an average $<c_2> = 2\times 10^{-3}$. The additive constant $<c_1>$ was fixed to zero as it was found to be consistent with zero. For this work, it was found that there is also a small dependence on the PSF strehl ratio $f_{\rm PSF}$ previously unaccounted for. In order to compute the new, non-parametric, $c$-correction, the PSF-corrected galaxy shapes $e_1$ and $e_2$ are binned in the three dimensional space $(\nu_{\rm SN}, r, f_{\rm PSF})$, where the statistical shape noise is roughly the same for each cell. In practice, the number of pixels in each dimension is not very important, we verified that the results are unchanged by dividing into $10^3$ or $30^3$ cells. The $c$-correction term is obtained by fitting a 3-dimensional third order polynomial for each component $e_1$ and $e_2$. The fitting procedure returns a $c_1$ and $c_2$ term as function of the bin position in the $(\nu_{\rm SN}, r, f_{\rm PSF})$ space, which are then assigned to each galaxy.  The new $c$-correction finds that both $c_1$ and $c_2$ are non-zero, although on average $c_1$ is of the order of $5\times 10^{-4}$, still a lot smaller than the average $c_2$ correction, which averages to $2\times10^{-3}$ as in \citet{2012MNRAS.427..146H}. The overall change on the cosmic shear signal between the previous and new $c$-correction is marginal (i.e. within the noise): the main difference however is  a change in the number of {\it bad} fields. Originally $42$ fields were flagged {\it bad}, while the new $c$-correction brings this number down to only $24$.
A further improvement is obtained when the field selection is performed on the same galaxies used for the analysis. Following \citet{2013MNRAS.433.3373V} we decided to restrict our analysis to the galaxies with photometric redshifts within $0.4<z_{\rm phot}<1.3$, where the number of $z_{\rm phot}$ outliers and the redshift errors are minimal. The final number of {\it bad} fields is $14$, yielding a total imaging area of $128$ deg. sq. of `good' data. The final step is to derive the redshift distribution $n(z)$ for the selected galaxies. As demonstrated in \citet{2013MNRAS.431.1547B}, the redshift distribution $n(z)$ is given by the {\it lens}fit-weighted stacked probability distribution functions of the galaxy sample $z_{\rm phot}$. In our case, the redshift distribution is well fitted with:
\begin{eqnarray}
  n(z) = N_0 e^{-(z-z_0)^4/\sigma_0^2}+N_1 e^{-(z-z_1)^4/\sigma_1^2} + N_2 e^{-(z-z_2)^4/\sigma_2^2}
  \label{eq:nz}
\end{eqnarray}
where  $(N_0,z_0,\sigma_0,N_1,z_1,\sigma_1,N_2,z_2,\sigma_2)$ = (0.54828, 0.69972, 0.07412, 0.59666, 0.81567, 0.21624, 0.20735, 1.1337, 0.30801). Fig. \ref{fig:fit_zPDF} shows the data and the best fit function. The mean redshift  between the two distributions differ by $1.2$ percent, which is well below the combined sources of error in our analysis. It is therefore neglected in the rest of the paper.


The shear correlation function measurement follows the same procedure as described in \citet{2013MNRAS.430.2200K}, by averaging over pairs of galaxies:
\begin{eqnarray}
\xi_{\pm}(\theta)={\sum_{i,j} w_i w_j \left[e_t(\theta_i) e_t(\theta_j) \pm e_r(\theta_i) e_r(\theta_j)\right]\over \sum_{i,j} w_i w_j}.
\end{eqnarray}
The sum is performed over all galaxy pairs $(i,j)$ with angular distance $|\theta_i-\theta_j|$ within some bin around $\theta$. The quantities $e_t$ and $e_r$ respectively denote the tangential and cross-component of the galaxy ellipticity. The weights $w_i$ are obtained from the {\it lens}fit shape measurement pipeline \citep{2013MNRAS.429.2858M}. This measurement is corrected by the shear calibration factor $1+K$ given by:

\begin{equation}
1+K(\theta)={\sum_{i,j} w_i w_j (1+m_i)(1+m_j)\over \sum_{i,j} w_i w_j}.
\end{equation}
The final calibrated measurements are obtained by dividing $\xi_{\pm}$ by $1+K$, which is $\sim0.89$ for all scales. The error on the calibration on the shear correlation function is completely negligible as shown in \citet{2013MNRAS.429.2858M}.

We also apply a conservative cut on the minimum angular separation for pairs of galaxies. \citet{2013MNRAS.430.2200K} used $9$ arc-seconds which corresponds to the image postage stamp analyzed by {\it lens}fit \citep{2013MNRAS.429.2858M} to measure galaxy shapes. We apply a cut at $20$ arc-seconds, which eliminates any possibility of the extended halo of a galaxy pair to be within the same fitted area. The measurement  uses the public code {\small ATHENA}\footnote{ATHENA: {\tt http://cosmostat.org/athena.html}},
and is  shown in Fig. \ref{fig:xi}. The results are divided by the fiducial DM-only model to present the differences between the data and the models. 
The inner and outer error bars show the statistical and combined statistical and sampling variance uncertainties, respectively.

\subsection{Simulations}
\label{subsec:sims}

This work makes use of the two SLICS simulation suites described in HDVW, which are based on WMAP9 + SN + BAO cosmology. 
The SLICS-LE suite consists of 500 independent $N$-body realization in which light cones of $60$ sq. degrees have been extracted
in the multiple thin lens and Born  approximations. It achieves better than 10 percent precision  for $\xi_+$ $(\theta > 0.4')$,
and down to the few arcminutes for $\xi_-$ $(\theta > 5')$. We use the LE suite to estimate the sampling variance component to the cosmic shear measurement.

The SLICS-HR series is a smaller ensemble of only five light cones in which the resolution is achieved for scales ten times smaller.  
It serves for convergence assessment and,  as mentioned in Section \ref{subsec:models}, as an independent estimate for  
the dark matter only $\xi_+$ signal. 
Details about the measurements of $\xi_{\pm}$ from these two simulation suites 
are provided in HDVW. 


\subsection{Theoretical predictions and measurements}

\subsubsection{Theoretical predictions}

\begin{figure*}
   \centering
   \hspace{-1cm}
   \includegraphics[width=6.0in]{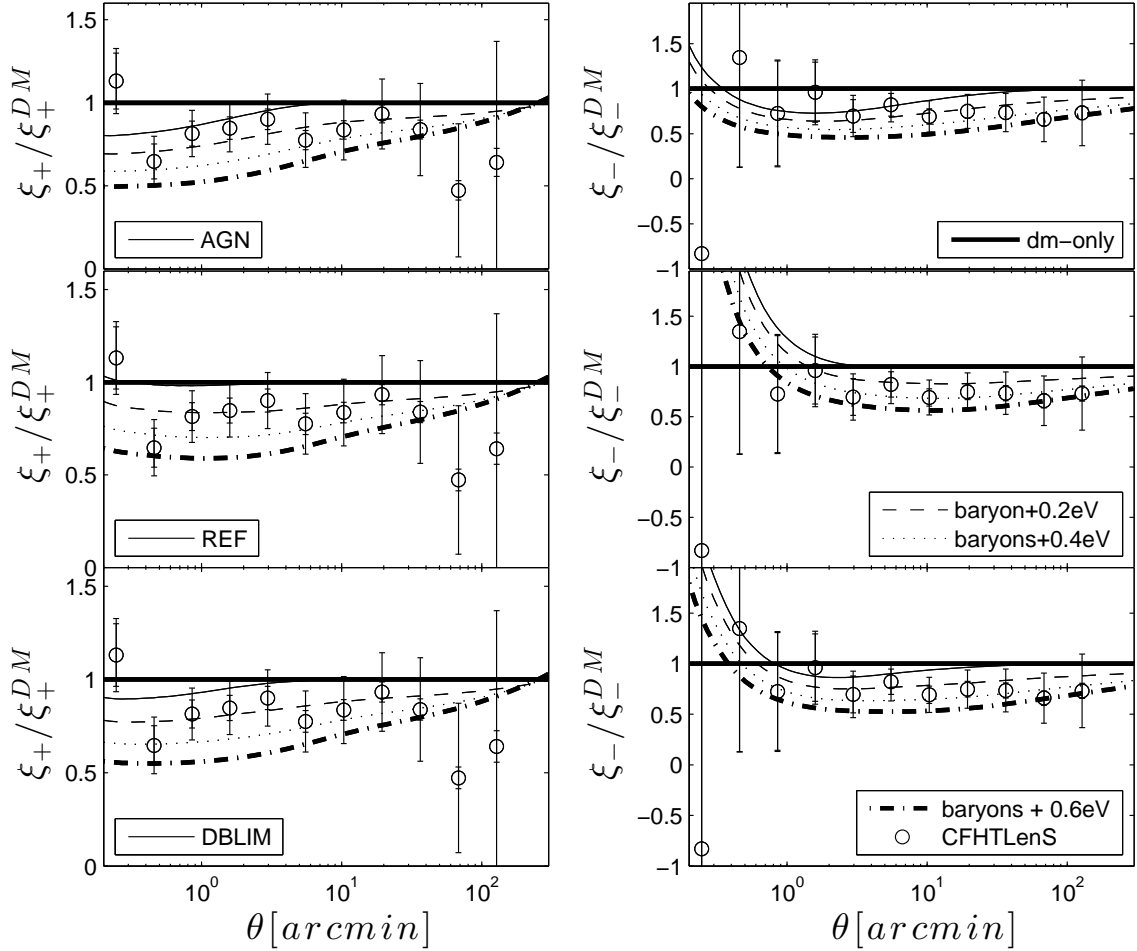} 
      \caption{Ratio between all predictions for $\xi_{\pm}$ and the dark matter only model from CEp. 
     Each panel shows the dark matter only model as the thick horizontal line, and the dark matter + baryon model as the thin solid line. 
      Top to bottom are the AGN, REF and DBLIM baryon models respectively. The impact of the neutrinos on each model is
      shown as the thin dashed lines ($0.2$eV), dotted lines ($0.4$eV), and the thick dash-dotted line ($0.6$eV).
      The open symbols are the measurements from  CFHTLenS, the inner error bars show the statistical error only, while the outer error bar combines
      all sources of error discussed in Section \ref{subsec:error}.}
   \label{fig:xi}
\end{figure*}

Fig. \ref{fig:xi} compares a range of model predictions  for the real space shear correlation functions $\xi_{\pm}(\theta)$ (obtained with equation \ref{eq:xi}),
with the measurements from the CFHTLenS data. As found by \citet{2011MNRAS.417.2020S}, we see that the baryonic feedback alone (thin solid line) tends to suppress the $\xi_+$ signal at small scales, with very little effect for scales  $\theta>5$ arcminutes, and that  
the maximum suppression ranges from zero to 20 percent, depending on the model.
The neutrino feedback (dashed, dotted and dot-dashed curves) adds an extra suppression that extends over a larger range of angles, exceeding 15 percent even at 100 arcminutes for $M_{\nu} \ge 0.4$eV. It is clear from these predictions that non-zero neutrino masses and baryon feedback have similar effects on the weak lensing power spectrum, both leading to power suppression of comparable magnitude.

The combined effect on $\xi_-$ is similar, except that the global shape is shifted to angles ten times larger; this is a simple geometrical effect due to the fact that, for the same angular separation $\theta$, $\xi_-(\theta)$ is probing smaller physical scales than $\xi_+(\theta)$.
 This could in principle allow for a sensitivity to the positive feedback of stars on the matter power spectrum, which occurs at $\theta < 1$ arcminute;
unfortunately this is also in a region where our measurements have the largest error bars, and hence cannot distinguish this feature.

 We see from Fig. \ref{fig:xi}  that both weak lensing shear estimators can be broken into two zones, separated at the scale where the baryonic feedback starts to have a significant  effect;
this occurs at $\theta = 5'$ and $40'$ for $\xi_+$ and $\xi_-$ respectively. The measurement from the `large angle' zone could serve to fix the neutrino mass with minimal contamination from the unknown baryon feedback mechanism, while the small angle zone could constrain (or include a marginalization over) the baryonic feedback model.
In this strategy, care must be taken to account for the high level of correlation that exists between the two zones, but this nevertheless could serve as a good starting point
for future weak lensing analysis. Note that the exact value of the zone separation angle will change with the source distribution $n(z)$.

We discuss the error bars in the next Section, however we can immediately see from their size that this data cannot distinguish a unique combination of baryon feedback model and neutrino mass.
However, certain combinations are unlikely and can even be ruled out with the current data set,  given our assumptions on the background cosmology are correct. In particular the dark matter only model seems already disfavoured.
Before detailing our model rejection technique (Section \ref{subsec:chi2}), we first describe our estimate of the full error that enters in this calculation, as this is a very important step for percent precision measurements.

\subsection{Error budget}
\label{subsec:error}

The total error in this measurement comes from the combination of statistical error, sampling variance, modelling error  and uncertainty in the background cosmology. 
Each of these contributions is discussed in this Section.

\subsubsection{Statistical } 

The shape noise generates a statistical error that dominates sampling variance at small scales. It is calculated from the measurement of $\xi_{\pm}$ in $200$ noise realizations, where the galaxy orientations extracted from the data have been randomized. The scatter in 
$\xi_{\pm}$ for each angular bin, and its covariance matrix across bins, is computed for all galaxy pairs that contribute to that particular bin. For the statistical noise, the covariance matrix is almost diagonal and the amplitude of the diagonal elements scale as $\theta^{-2}$.

\subsubsection{Sampling variance} 

\begin{figure}
   \centering
   \includegraphics[width=2.5in]{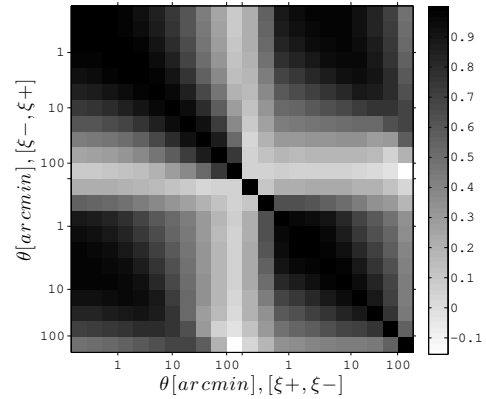} 
      \caption{Cross-correlation coefficients about $\xi_+$ (top left block),  $\xi_-$ (bottom right) and the symmetric cross terms (top right and bottom left),
       measured from the large ensemble of 500 simulations. We recover from  the two off-diagonal blocks that the measurements of $\xi_-$ 
       at 10 arcminutes correlate very strongly with those of $\xi_+$ at 1 arcminute.}
   \label{fig:r_xi}
\end{figure}

The sampling variance is estimated from the LE simulation suite by computing the quantity $\mbox{Cov}^{\xi\pm \xi\pm}_{N-body}(\theta, \theta')$.
We calculate this quantity for the two auto-correlation $(++, --)$ terms plus the  cross-term $(+-)$ in preparation for the combined analysis (see Fig. \ref{fig:r_xi}).
Since the covariance is inversely proportional to the area,  we rescale each of these three quantities by the ratio of the
simulation light cones and the CFHTLenS unmasked areas spanned by the good fields (i.e. 60/128) in order to match the sky coverage of the data. 
We correct for the finite support effect described in HDVW, although this has a sub-percent impact on the sub degree scales under study.

We also include the mixed term arising from the coupling between the shot noise and the sampling variance. 
We follow the results from  \citet{2013MNRAS.430.2200K} in that the mixed term closely follows the sampling variance term, aside from an overall normalization term
taken to be $0.25$ and $1.0$ for $\xi_+$ and $\xi_-$ respectively.  We verified for the case of $\xi_+$ that this closely reproduces the analytical calculations described in \citet{2002A&A...396....1S}, where the effective galaxy densities $n_{\rm eff}$ and the dispersion in the measured galaxy ellipticities $\sigma_{\epsilon} $ are 
taken to be $n_{\rm eff} = 9.2$ gal. arcmin$^{-2}$ and $\sigma_{\epsilon} = \sqrt{\sigma^2_{e1} + \sigma^2_{e2} } = 0.395$, respectively. We report this analytical calculation   in Fig. \ref{fig:budget} (dotted lines).

We correct for the finite mass resolution in the simulations, a limitation that results in a lack of structure at small scales, causing a drop in both the signal and the covariance. 
This {\it missing power}  can be quantified by comparisons against reliable prediction models or higher resolution simulations, the SLICS-HR series in this case.
The actual impact on the covariance matrix can be estimated from the Hyper Extended Perturbation Theory \citep{1999ApJ...520...35S}, which states that the covariance in power spectrum scales as $\mbox{Cov}(k,k) \propto P^{3}(k)$ in the non-linear regime. We therefore use this scaling relation to correct the covariance about $P(k)$, keeping the off-diagonal cross-correlation coefficients fixed, and propagate 
the effect onto the weak lensing covariance matrix using the Limber approximation (Harnois-D\'eraps \& van Waerbeke in prep.). We show the impact of this correction in Fig. \ref{fig:budget} as the red-dashed line. The largest effect is a 10 percent and 200 percent  increase in the error about $\xi_+$ and $\xi_-$ respectively at $\theta<0.5'$.
Above 1 arcmin $(\xi_+)$ and 10 arcmin $(\xi_-)$, the correction is negligible.

 The baryonic feedback and neutrino free streaming both suppress the small scale power, which can to some extent be matched to the 
power loss in the simulations due to mass resolution limits in the $N$-body calculation. One could then argue that if neutrinos are massive and/or baryon feedback suppresses the matter power in the real Universe, then the mass resolution correction most likely overestimates the error, and  the sampling variance computed without this correction  is more accurate.
This is a valid concern, and should be investigated in the context of future surveys, ideally correcting the covariance in a manner consistent with the model under study. For this work, however, we stay conservative and apply the same correction to all models, keeping in mind that the sampling variance in the case of massive neutrinos and baryon feedback 
model will be slightly overestimated and hence 
our constraining power at rejecting these models is slightly too weak.

\subsubsection{Non-linear modelling} 

 As discussed  in Section \ref{subsec:models} and observed in Fig. \ref{fig:model_err}, the scatter between the predictions from different pure dark matter non-linear models is 
  always less than five percent for $\xi_+$, while it exceeds 50 percent in the smallest angles of $\xi_-$. 
 Different models and different angles have been weighted by their inverse variance, and 
 the resulting weighted errors on the non-linear  models are no more than 4 and 8 percent for $\xi_+$ and $\xi_-$ respectively (see Section \ref{subsec:models} for more details).
 This is treated as a source of systematic uncertainty in our calculation. Comparing this to the  other sources of error in Fig. \ref{fig:budget},
 we find this error to be sub-dominant.

 \subsubsection{Cosmology} 
 \label{subsubsec:cosmo_err}
 
 In this analysis, we fix the background cosmology to that found by WMAP9+BAO+SN assuming a flat $\Lambda$CDM cosmology, 
in order to probe the impact of neutrinos and baryon  feedback.  Weak lensing is very  sensitive to both the amplitude of the matter power spectrum, characterized either by $\sigma_8$ or $A_s$, and the matter density parameter, $\Omega_M$, with the shear correlation functions scaling roughly  as $\xi_{\pm} \propto \sigma_8 \Omega_M^2 \propto A_s^2 \Omega_M^2$.
The WMAP9 constraints on $A_s$ are precise to 3.3 percent, and on $\Omega_M$ to 3.4 percent  \citep{2013ApJS..208...19H}, 
and we factor these uncertainties into our analysis through an additional error in our systematic error budget (see dashed, thin lines in Fig. \ref{fig:budget}).
Comparing this `cosmological' uncertainty to the other sources of error in Fig. \ref{fig:budget}, we find it to be sub-dominant compared to the statistical shot noise (solid, thick) and 
sampling variance (dashed, thick), as expected from a comparison of cosmological constraint from CFHTLenS data alone \citep[e.g.][]{2013MNRAS.430.2200K} with WMAP9. It is however more significant than the uncertainty on the non-linear modelling of the dark matter only signal.

 The combined (non-linear model + cosmology) uncertainty on $\xi_{\pm}$ is shown as the error bars about the open circles in Fig. \ref{fig:model_err}.
We observe that on small angular scales, both contributions are of the same magnitude, whilst the cosmology errors at large angles are dominant.

\subsubsection{Halo sampling variance}

Another source of error on the measurement -- from both data and simulations -- comes from the {\it halo sampling variance}  (HSV hereafter),
which is caused by the finiteness of the observation volume. This effect has been studied in terms of the halo model by \citet{2009ApJ...701...945S},
which has shown that it can be described by an extra term in the covariance matrix in multipole space: 
\begin{eqnarray}
\mbox{Cov}_{HSV}(\ell, \ell') = \bar{b^2} \sigma^{2}_{RMS}(\Theta_s)C^{\kappa,1h}_{\ell}C^{\kappa,1h}_{\ell'}
\end{eqnarray}
where $\bar{b^2}$ is the mean halo bias, $ \sigma_{RMS}(\Theta_s)$  is the RMS fluctuations in angular clustering inside a circle of area $A$ and radius $\Theta_s = \sqrt{A/\pi}$,
and $C^{\kappa,1h}_{\ell}$ is the one-halo contribution to the lensing power spectrum, averaged over all halo masses.
We propagate this quantity onto our real space weak lensing estimators $\xi_{\pm}$ as in \citet{2008A&A...477...43J}, i.e. using a two-dimensional equivalent of equation \ref{eq:xi} and converting $\mbox{Cov}_{HSV}(\ell, \ell')$ into $\mbox{Cov}_{HSV}^{\xi\pm}(\theta, \theta')$. We show the contribution to the covariance coming from the HSV in Fig. \ref{fig:budget} (dot-dashed), 
and observe that it is subdominant everywhere. When added in quadrature, it would contribute less than a percent to the total error, hence it can be safely ignored.

 \subsubsection{Error on the sampling covariance} 

The residual error in estimates of sampling variance derived from $N$-body simulations propagate as an 
{\it extra error} on the cosmological parameters \citep{2007A&A...464..399H, 2013PRD.88.063537D}. The size of this error scales as $1 + N_{data}/N_{sim}$, i.e. the ratio
between the size of the data vector and the number of independent simulations that enter the estimate. In our case, the full data vector ($\xi_+$ and $\xi_-$ combined) consists of
22 elements, which, when divided by $N_{sim} = 500$, would contribute a 4 percent error -- and two percent for the $\xi_+$ only measurement -- on the precision of 
cosmological parameters derived from the cosmic shear data. 
In this analysis, we do not search for cosmological parameters, but instead perform a hypothesis rejection procedure,
which is  less sensitive to  this {\it extra error} and  can therefore be neglected. 

 \subsubsection{Other potential sources of error} 

The interpretation of the weak lensing signal is in many cases blurred by contamination from secondary effects,
the most dominant being the intrinsic alignment that exists between galaxies that are tidally connected. 
This becomes highly important for analyses based on tomography \citep{2013MNRAS.432.2433H} or full three-dimensional lensing \citep{2014MNRAS.442.1326K}, 
but is a  weak effect in our case, owing to the full collapse of the survey along the radial coordinate.  
We therefore do not include an intrinsic alignment modelling error in our uncertainty.
The random error from shape measurements is already absorbed in the statistical error 
and are therefore not contributing as separate terms. The error on photometric measurement would affect the modelling of the signal 
via an incorrect estimate of $n(z)$, which would affect the amplitude of the signal. 
However the uncertainty on this quantity is much smaller than our `cosmological' error and is therefore not included.

\subsubsection{The budget}

\begin{figure}
   \centering
    \includegraphics[width=2.7in]{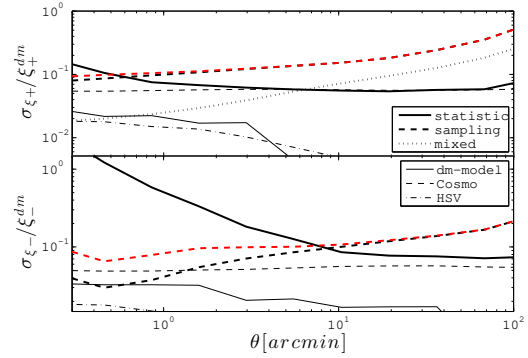} 
      \caption{A comparison of the different contributions to the error budget on CFHTLenS  measurements of $\xi_+$ (upper) and $\xi_-$ (lower) as a function of scale.
        The errors are  dominated by statistical shot noise (solid, thick) on small scales and sampling variance (dashed, thick) on large scales. The impact of
        re-scaling the small angle signal  to account for mass resolution effects can be seen by comparing the sampling variance with (red, top) and without (black, bottom)
        this scaling. The mixed term is shown by the dotted lines (for $\xi_+$ only) and is smaller than the sampling variance at all scales by at least a factor of 2. 
        For $\xi_-$, it is taken to be identical to the sampling variance, in good agreement with \citet{2013MNRAS.430.2200K}.
        By fixing the background  cosmology, we include a secondary level error term (dashed, thin) which includes a $3.4$ percent uncertainty on $\Omega_M$
        and a $3.3$ percent uncertainty on $A_s$.
        The error arising from our uncertainty on the dark matter only non-linear model (solid, thin) is an order of magnitude smaller that the largest error and hence negligible in this measurement. 
      The HSV term (dot-dashed) is sub-dominant everywhere and can be safely ignored.}
   \label{fig:budget}
\end{figure}

We show the different error contributions to the shear correlation measurements in Fig. \ref{fig:budget}. 
The error budget on $\xi+$ is dominated by the sampling variance above one  arcminute, and by statistical uncertainty at smaller angles. 
For $\xi_-$, the statistical error dominates to  scales $\theta<10$ arcminutes with sampling variance dominating at larger scales.
Our uncertainty on the background cosmology is at most 50 percent of the sampling variance,
while the non-linear model uncertainty is sub-dominant at all scales.
For both weak lensing quantities, the off-diagonal elements of the covariance are completely dominated by the sampling variance, 
since the random noise is highly diagonal.

\subsection{Model rejection}
\label{subsec:chi2}
 
As discussed by \citet{2014arXiv1405.6205N}, the effects of a non-zero neutrino mass are degenerate with  baryonic feedback, particularly at small angular scales.
Varying the DM parameters $A_s$ and $\Omega_M$  also changes the model on these scales, hence future weak lensing 
analyses will need to carefully address these degeneracies in the parameter estimations.
Although an MCMC analysis would give a complete story, we take here a first step proceeding with a 
case-by-case model rejection,  based on a measurement of the $\chi^2$ for each model and, finally, of the $p$-value. 
The $p$-value captures the statistical significance of the measurement, or, in other words, probability that the data is consistent with the model, {\it if the model is true}. 
It is simply given by the integral of the $\chi^2$ probability density function, from the measured $\chi^2$ up to infinity. Lower values represent higher levels of model rejection. Following standard statistics, $p$-values of (0.317, 0.046, 0.003, ...) correspond  to model rejection at the $(1\sigma, 2\sigma, 3\sigma,...)$ level.
This choice of discrimination strategy is driven by the fact that the baryon feedback models are not described by a continuous parameter, 
meaning that we cannot perform a full likelihood fit to extract a set of baryonic feedback models out of a smooth distribution\footnote{
What we could extract from a MCMC analysis are the preferred values for the parameters of Table \ref{table:fit_baryon}, but these 
then need to reconnect with the feedback models, which ultimately resemble the analysis we present in this paper.}.
Each model is unique and has to be tested independently against the data.

\begin{table*}
   \centering
   \caption{Distribution of $p$-values for different combination of baryon feedback models and neutrino masses (see main text for details). 
   Specifically, each entry in this Table represents the largest $p$-value probed inside a $3\sigma_{syst}$ region about the mean of the model. 
   Values in bold face highlight the model combinations that are excluded by the data with more than $1.64 \sigma$ significance ($p$-value $<0.1$, equivalent to a confidence interval (CI) of 90\%).  }
      \begin{tabular}{@{} ll|l|ccccccccc @{}c} 
      \hline
             &       &  & \multicolumn{4}{|c|}{$\xi_+$ alone} & &  \multicolumn{4}{|c|}{$\xi_+$ and $\xi_-$ combined}\\
             \hline
             &         $M_{\nu}$& & 0.0eV              & 0.2eV & 0.4eV & 0.6eV & &  0.0eV              & 0.2eV & 0.4eV & 0.6eV \\
      \hline
 & DM-only  && \bf0.036 & 0.158 & 0.267 & 0.230 && \bf0.037 & 0.278 & 0.621 & 0.760\\
 & AGN      && 0.110 & 0.222 & 0.209 & 0.120  && 0.168 & 0.476 & 0.659 & 0.675\\
 & REF      && \bf0.050 & 0.189 & 0.289 & 0.235 && \bf0.030 & 0.234 & 0.571 & 0.728\\
 & DBLIM    && \bf0.092 & 0.245 & 0.283 & 0.191 && 0.109 & 0.438 & 0.695 & 0.756\\

          \hline
      \end{tabular}
    \label{table:pvalues}
\end{table*}

 Since we are testing individual models, as opposed to performing a thorough MCMC calculation,  it is important to adopt a strategy to account for the three sources of systematic 
uncertainty -- i.e. that on $A_s$,  $\Omega_M$ and on the non-linear dark matter only model.  
We proceed as follows : for each combination of neutrino mass and feedback model, 
we allow the amplitude of the data signal to vary within a 3$\sigma_{syst}$ range about the measured value and search   
for the most favourable hypothesis (highest $p$-value). The systematic uncertainty 
is maximal at small angles and reaches up  to $\sim9$ per cent of the model amplitude for both $\xi_+$ and $\xi_-$. This means that for our calculation of the $p$-value, 
we allow the data points to shift up and down by up to $27\%$  on Fig. \ref{fig:xi},
keeping the shown  error bars (statistical + sampling) fixed. 
Given that the Planck value for $\Omega_m$ is roughly $2\sigma$ higher than the WMAP9 best measurement, 
our $3\sigma_{syst}$ excursion allows for an nice overlap between both data sets.

Statistically, our model rejection method is equivalent to fitting $(A_s^2\Omega_M^{1.8})$ from the amplitude of the cosmic shear signal, then estimating the neutrino mass for each baryon feedback model from the largest $p$-value, although our sampling in the $M_{\nu}$ direction has only four points.
Accordingly, the number of degrees of freedom must reduced by two in the conversion
between $\chi^2$ and $p$-values.


We consider two cases, one where the data vector only includes $\xi_+$ and one with both $\xi_+$ and $\xi_-$. The resulting $p$-values are summarized for all our results in Table \ref{table:pvalues}. The models rejected at more than $1.64\sigma$ (i.e. 90\% CI) are highlighted in bold,. 
We see that for most $\xi_+$ models, the $p$-value is the highest in the  $0.4$eV column and the smallest in the $0.0$eV column, indicating that the preferred value for $M_{\nu}$ is non-zero, although a zero neutrino mass cannot be ruled out. 
The combined $\xi_{\pm}$ measurement seems to prefer even higher values of $M_{\nu}$, but the significance of this statement is weak given the size of the $p$-values.
The dark matter-only model (zero neutrino mass and no baryonic feedback) is however rejected with more than  $2\sigma$. 
It is clear from Fig. \ref{fig:xi} that the discriminating power is maximal in the region $\theta < 10 $ arcminutes.
 The combination (REF and massless neutrino) is also excluded at more than $2\sigma$ by the data, for having too much power at small angular scales;  
at the same time, the combination (DBLIM and $M_{\nu}=0.0$eV) is rejected with 91\% confidence.
Some of the scenarios with massive neutrinos considered here are disfavoured by the $\xi_+$ data but only with weak significance ($88\%$ confidence for the rejection of the AGN feedback model combined with $M_{\nu}=0.6$eV).

Since the first angular bin at $\theta=0.23$ arcminute is in tension with the models compared to the other bins, 
as a sanity check, we explored the impact of excluding that data point, finding no changes on our conclusions, only a modest reduction in the statistical constraining power. 
This first bin is stable against different minimum separation cuts. As explained in Section 3.2.2, the minimum cut at $\theta_{min}=20$ arcsec is very conservative to guarantee that two galaxies do not fall within the same {\it lens}fit template. In \citet{2013MNRAS.429.2858M}, image simulations showed that the shape of close galaxies were not biased even down to $5$ arcseconds separation. We tried various cuts $\theta_{min}$ from $5$ to $20$ arcseconds, and the results shown in Fig. \ref{fig:xi} do not change, we are therefore confident that the location of the first bin is robust and not the result of unaccounted residual systematics.
 
 \section{Discussion and Conclusion}
\label{sec:discussion}

In this paper, we have considered the use of weak gravitational lensing to probe baryonic feedback and neutrino masses through their effect on the mass power spectrum. 
For this purpose, we constructed a fitting formula that describes the effect of baryons on the mass power spectrum for three specific models studied in \citet{2014MNRAS.440.2997V}. This formula is an analytic function of redshift $z$ and physical scale $k$, therefore it can be used in cosmological forecasting and MCMC chains, even in the non-linear regime. Our fitting function is highly accurate over the redshift range $0<z<1.5$ and scales $k<100 h/\mbox{Mpc}$, and can be extended to $z=3$ with a modest degradation in precision at the smallest scales. It can be used for a wide range of cosmological applications, including comparisons between different sets of hydrodynamical simulations, or even high precision baryonic acoustic oscillations measurements \citep{2014MNRAS.442.2131A}. 

This formula was used to make predictions for the CFHTLenS weak lensing data. We find that the data, in combination with WMAP9 cosmological parameter constraints, 
reject with {\it at least} 90\% confidence  1) the dark matter only model  (i.e. massless neutrinos and no baryon feedback, $96\%$ confidence), 
2) the combination of massless neutrinos with baryon feedback model REF ($97\%$ confidence),  and 3) the combination of baryon feedback model DBLIM with massless neutrinos ($91\%$ confidence). These are strong hints that neutrinos are indeed massive, although the massless scenario cannot be completely ruled out in this analysis. 
The data also disfavour other combinations, although with a lower significance. 

Future weak lensing surveys with larger total area will be very promising for this type of analysis, since the CFHTLenS error budget is currently dominated by sampling variance and statistical error. The completed RCS2 survey with its re-analysis RCSLenS\footnote{RCSLenS: {\tt www.rcslens.org}} covers close to 700 deg$^2$, the ongoing KiDS and HSC  will cover 1500 deg$^2$ each, 
while DES will cover more than 5000 deg$^2$. These data sets combined represent a sky area that is $\sim60$ times larger than the CFHTLenS survey considered in our study. 
We show in this paper how we can use the intermediate angle region $(\theta > 5'$ and $40'$ for $\xi_+$ and $\xi_-$, respectively) to fix the neutrino mass, then examine and constrain the baryon feedback models with the smaller angles.

Future lensing studies could also probe feedback models as function of galaxy type, age or environment, and study how the density profile of the dark matter halo is affected in a non-uniform manner \citep{2014MNRAS.442.2641V,2014arXiv1406.5013F}. Galaxy-galaxy lensing in particular is a promising area where these ideas could be implemented. Cross-correlation studies that are sensitive to feedback effects, such as the cross-correlation between thermal Sunyaev-Zeldovich and gravitational lensing \citep{2014PhRvD..89b3508V,2014arXiv1404.4808M}, are also particularly ideal for constraining these models.

 It will be interesting to explore whether a tomographic study could help disentangling baryonic feedback from massive neutrinos. One should be careful in that case, however, to take into account intrinsic galaxy alignment, as it is a non-negligible correction to the lensing signal for three redshift bins or more \citep[see][for example]{2013MNRAS.432.2433H}.

Different feedback models are currently given by specific hydrodynamical simulations, but one can envision a not so distant future where it will be possible to implement galactic feedback as just another set of parameters to be simultaneously fit  with other cosmological parameters.

\section*{Acknowledgements}

 We are deeply grateful to Lance Miller for his essential work on the {\it good/bad} field selection code and shear calibrations pipelines, and to all the CFHTLenS team for having made public their high quality shear data. We also thank Fergus Simpson and Martin Kilbinger for providing their comments on the draft.
Computations for the $N$-body simulations were performed on the GPC supercomputer at the SciNet HPC Consortium. 
SciNet is funded by: the Canada Foundation for Innovation under the auspices of Compute Canada; 
the Government of Ontario; Ontario Research Fund - Research Excellence; and the University of Toronto. 
JHD is supported by a CITA National Fellowship and NSERC, and LvW is funded by the NSERC and Canadian Institute for Advanced Research CIfAR.
MV is funded by grant 614.001.103 from the Netherlands Organisation for Scientific Research (NWO). 
MV and CH acknowledge  support from the European Research Council under FP7 grant number 279396 (MV) and 240185 (CH). This work is based on observations obtained with MegaPrime/MegaCam, a joint project of CFHT and CEA/IRFU, at the Canada-France-Hawaii Telescope (CFHT) which is operated by the National Research Council (NRC) of Canada, the Institut National des Sciences de l'Univers of the Centre National de la Recherche Scientifique (CNRS) of France, and the University of Hawaii. This research used the facilities of the Canadian Astronomy Data Centre operated by the National Research Council of Canada with the support of the Canadian Space Agency. CFHTLenS data processing was made possible thanks to significant computing support from the NSERC Research Tools and Instruments grant program.





\bibliographystyle{hapj}
\bibliography{mybib3_new}

\bsp

\label{lastpage}

\end{document}